# How Platform Exchange and Safeguards Matter: The Case of Sexual Risk in Airbnb and Couchsurfing

SKYLER WANG, Sociology, University of California, Berkeley, USA

Recent work in CHI and CSCW has devoted increasing attention to how the design of network hospitality platforms shapes user experiences and relational outcomes. In this article, I interrogate how different risk factors emerge based on the type of exchanges these platforms facilitate. To do so, I juxtapose two prominent network hospitality platforms—one facilitating negotiated exchange (i.e., Airbnb) with another facilitating reciprocal exchange (i.e., Couchsurfing). Homing in on *sexual* risk, an underexplored form of platform danger, and drawing on interviews with 40 female dual-platform users, I argue that Airbnb's provision of binding negotiated exchange and institutional safeguards reduces risk through three mechanisms: *casting initial guest-host relation into a buyer-seller arrangement*, *stabilizing interactional scripts*, and *formalizing sexual violence recourse*. Conversely, Couchsurfing's focus on reciprocal exchange and lack of safeguards increase sexual precarity for users both on- and off-platform. This study demonstrates how platforms with strong prosocial motivations can jeopardize sociality and concludes with design implications for protecting vulnerable user populations.

## 1 INTRODUCTION

As part of the sharing economy, network hospitality platforms such as Airbnb and Couchsurfing leverage the logic of peer-to-peer exchange to provide access to "otherwise underutilized" apartments or homes [23,24,35:531]. By enabling strangers to match through an online network and arrange offline interactions, these platforms gave rise to a form of sociality centered around the concept of shared hospitality [23]. While they promulgate moralistic goals such as advancing sustainable tourism, cross-cultural contact, and the revitalization of homestay culture [22], recent scholarship has identified a noticeable prevalence of user risks (e.g., crime, robbery, surveillance, etc.) within these communities [59,68]. In this article, I build on previous work in Computer-Human Interaction (CHI) and Computer-Supported Cooperative Work (CSCW) concerning network hospitality [29,31–33,37] and platform-mediated violence

Author's address: Skyler Wang, skyler.wang@berkeley.edu; University of California, Berkeley, CA, USA.

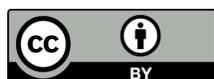

[53,61,72] to advance our knowledge of an underexplored type of risk that has garnered increasing media attention in recent times [9,71]—those that are *sexual* in nature [8]. While these risks can lead to various forms of sexual violence (i.e., activities of a sexual nature sans consent), they reveal themselves as systemically relevant when hosts and guests co-mingle within the intimate spaces of a stranger's home.

Beyond modes of exchange, I further interrogate how the provision of institutional safeguards by platforms—impersonal mechanisms that give confidence to those engaging in exchange [36,38]—plays a role in shaping sexual risk. As described by Lampinen and Cheshire [37], an example of such an institutional safeguard involves third-party platforms like Airbnb *binding* the negotiated exchange between guests and hosts. Doing so serves two primary functions: first, it lends structure to hosts' and guests' core exchange, reduces interactional risk, and may even catalyze prosocial interactions such as gifting and sharing. Next, the financial gains commercial platforms derive from binding these transactions allow them to direct resources into designing and implementing safety features, customer support programs, and user guarantees that act as a safety net encouraging users to enter into exchange relationships with strangers. Here, I build on Lampinen and Cheshire's [37] work to underscore specific mechanisms and conditions that make a platform the 'third actor' in an exchange loop, and how this triadic arrangement (guest-host-platform) affects users' perception of and experience with sexual risk both on- and off-platform.

Methods-wise, this study leverages interviews with 40 *female* Airbnb and Couchsurfing users—a group disproportionately affected by sexual violence in network hospitality [12]—to paint a less portrayed aspect of network hospitality. Even though many participants occupy both the role of host and guest (especially in the context of Couchsurfing), I focus specifically on their experiences as *guests* to counterbalance the bias towards spotlighting hosting experiences in previous network hospitality studies (for examples, see [29,30,36,37]). In addition, although sexual violence is prevalent on both platforms, I demonstrate across three stages—choosing who to stay with, in-person interactions, and responses to sexual harassment or violence—that participants overwhelmingly regard Airbnb as the sexually safer platform. By binding negotiated exchange and providing institutional safeguards, Airbnb reduces risk through three mechanisms: *casting initial guest-host relations into a buyer-seller arrangement*, *stabilizing interactional scripts*, and *formalizing sexual violence recourse*. Conversely, Couchsurfing's facilitation of reciprocal exchange and general lack of safeguards elevate various forms of sexual risks both on- and off-platform. More specifically, Couchsurfers experience greater degrees of sexual pressure and are less likely to report sexual violence incidents to the platform. In sum, this study makes a valuable contribution by empirically demonstrating how platforms with strong social motivations can, at times, produce higher levels of precarity. I conclude with implications for design and future research that could make network hospitality and the sharing economy safer for certain vulnerable user populations.

## 2 BACKGROUND

In this section, I first review human-computer interaction (HCI) literature related to platform-mediated sexual risk. Then, I connect this branch of research to social exchange theory within the context of network hospitality. Subsequently, I identify and explore how we can better understand sexual risk via the lens of exchange and safeguards using the case studies of Airbnb and Couchsurfing.

## 2.1 Platform-mediated Sexual Risk

Much ink in HCI research has been spilled on elucidating the Janus-faced nature of digital sociality—that while it enables new possibilities, it concomitantly exposes participants to various risks and harms. Within this purview, sexual risk generally refers to the potential dangers that emerge from digitally mediated communications or interactions and encompasses a range of experiences [27]. HCI work in this area has examined issues pertaining to online sexual harassment [47], computation approaches to detecting online sexual risk [50,51], the impact of sexual risk on youths and adolescents [1,17], and how such risk factors transpire in emerging forms of virtual reality environments [6,57].

Importantly, sexual risks do not stay within the confines of digital channels. For platforms that facilitate online-to-offline sociality, such risks begin online and are subsequently transferred to physical settings. That means beyond emotional or psychological harm, sexual violence arising from these in-person interactions could lead to physical abuse and assaults. As demonstrated by Zytko et al.'s [72] study of sexual consent on Tinder, sexual risk negotiation cuts across online and offline modalities, and platforms ought to consider holistic experiences when devising strategies that prevent systemic occurrences of sexual violence. Other scholars have also examined how platform-mediated sexual risks intersect with problems commonly associated with public health and disease management [62,67].

## 2.2 Social Exchange in Network Hospitality

Although Klein et al. [33] alluded to network hospitality's sexual violence problem in their work, this article marks the first attempt at investigating this issue systematically. Network hospitality, by definition, comprises online-to-offline systems that bring strangers together in the physical world to engage in peer-to-peer exchange [23]. Aside from exchanging material resources (i.e., a couch or living space), immaterial resources—which include language or cultural exchange—can be just as germane to one's experience within this space.

That said, *how* these resources are exchanged is as important as what is being exchanged. In this study, I direct attention to two primary forms of social exchange (i.e., how "a series of sequential transactions take place between two or more parties" [16:1])—negotiated and reciprocal. *Negotiated* exchanges, which are common in commercial settings, involve actors jointly negotiating the terms of a contractual agreement [44]. While not necessary, money is often involved in these transactions, and agreements are typically bound by a third party. Social exchange on paid platforms is typically predicated on this binding negotiated archetype [13]. As a negotiated form of exchange, guests of these platforms search for accommodation that fits their travel needs and budget. Upon payment, platforms—acting as an intermediary and "assurance structure"—confirm and bind the booking by formalizing the transaction between payee and payer [37]. As a transaction mode, binding negotiated exchange reduces uncertainty between exchange partners (since they can rely on assurance structures to enforce a transaction), which has direct implications on sexual risk, a point I will return to below.

Next, there is *reciprocal* exchange. In contrast to binding negotiated exchange, reciprocal exchange typically involves one actor giving something to another without knowing "whether, when, or to what extent the other will reciprocate in the future" [44:3]. Unlike paid platforms, where interested users directly pay for their desired place, those using reciprocal-based systems typically search through a selection of hosts' profiles and send private messages to express interest. Guests and hosts then use initial online conversations to learn about each other and gather interest, but explicit discussions of what a guest ought to offer in return are taboo. This mode of exchange underpins the logic of *gifting*. As Spence [58] argues, gifts are marked by a deep sense of inalienability—in addition to embodying a giver's identity and interests, they conjure a bond-affirming effect. By offering their accommodation, hosts invite a relationship by extending the opening gift of hospitality to their guests. But, as Mauss' [41] canonical work on gifting suggests, gifts are fundamentally motivated by three obligations—to give, receive, and *reciprocate*. While a gift obfuscates any economic intentions a giver may have, it nevertheless compels a form of reciprocal obligation in its recipient. Accordingly, indebted guests, who may choose to counter-gift through various forms of cultural exchange, could also become more vulnerable to sexual pressures [15]. At times, these dynamics could turn into sexual coercion, increasing the risks involved for guests.

## 2.3 A Case Study of Airbnb and Couchsurfing

Below, we dive deeper into the specific platforms chosen for this study and how sexual risks intersect with the different modes of exchange and safeguards they promote.

*2.3.1 Airbnb: Voluntary Sociality in a Binding, Negotiated Platform.* Having gone public in 2020 and with a valuation of over $100 billion, Airbnb has over 150 million users in 191 countries worldwide today [28]. Despite not owning any of its listed properties, Airbnb brings in an annual revenue of over three billion dollars [74], making it one of the world's most profitable hospitality corporations. The platform generates most of its revenue from service fees charged to guests and hosts with every booking. Guests, who bear the bulk of such costs, are generally required to pay up to an additional 14.2% of the total booking subtotal to complete a transaction. In comparison, hosts are charged a 3% fee with each booking.

As mentioned in the introduction, Airbnb is a platform that facilitates binding negotiated exchanges. In doing so, Cheshire et al. [13:178] argue that Airbnb reduces risk and uncertainty for hosts insofar as the core exchange is "guaranteed by law or a third-party mediator." More specifically, the fee Airbnb collects with every transaction is partly earmarked for financial payouts issued to hosts in the event of property damage. While it appears that such schemes primarily protect the interests of hosts, guests stand to gain from them as well. In fact, Airbnb's rebranded consumer protection policy, AirCover (introduced in May 2022), reifies protections for *guests*. For instance, if a host makes a guest feel sexually uncomfortable, guests can contact Airbnb's 24/7 customer support to request a refund or re-accommodation. Taken collectively, these safeguards may instill a sense of institutional trust, positioning Airbnb as a critical structure that undergirds trust-based relations on the platform [2]. For an enterprise that relies on "stranger sharing" [55], binding negotiated exchange and institutional safeguards, alongside Airbnb's reputation-based review system [19], may bolster user confidence in renting homes from individuals with whom they have minimal prior connections.

Although it may appear that guest-host interactions on Airbnb are becoming increasingly sparse given the swelling popularity of entire home rentals, self-check-ins, and external property management companies, a 2017 Boston study [43] found that half of the listed properties on Airbnb in the city are private (47.6%) or shared rooms (2.89%). This statistic, alongside budgetary limitations shared by many travelers, suggests that in many places, guests continue to interact with their hosts physically in one capacity or another. The intimacy of sharing a living area and the connection these exchanges may ensue suggest interactions of a sexual nature are not out of the question either. That said, given that the social obligation for paid guests and their hosts to connect is low [29], guests may have greater control over how intimate exchange transpires. Such control may, however, be complicated by certain gender pairings (for example, when a female guest stays with a male host [33]).

*2.3.2 Couchsurfing: Social Obligations in a Reciprocity-based Platform.* Couchsurfing comprises a global community of 14 million users and fetches an annual revenue of approximately four million dollars [75]. Unlike Airbnb, Couchsurfing does not profit from facilitating financial transactions between guests and hosts. In its earlier days, Couchsurfing largely depended on charitable donations and voluntary member verification fees to keep the platform afloat [46]. After being denied non-profit status in 2011, Couchsurfing transitioned from its free-to-use model to a 'freemium' configuration where users are encouraged to pay to verify their accounts and access exclusive features. The success of this verification model, however, is limited; as of 2020, only 4% of active members have verified profiles [73]. During the recent COVID-19 pandemic and on the brink of shutting down due to dwindling user traffic, the platform mandated a $2.39 monthly or $14.29 annual fee for all members in May 2020. Despite these changes, Couchsurfing continues to pride itself on facilitating hospitality trading between users without the exchange of money.

In contrast to the binding negotiated exchanges found in Airbnb, Couchsurfers participate in reciprocal exchanges [13,25,41]. Contrary to the voluntary nature of Airbnb "hangouts," guest-host interactions in Couchsurfing are markedly more obligatory [33]. Explicit reciprocity, including cooking hosts a meal or leaving a positive reference, and implicit reciprocity, including working with hosts' schedules to ensure guests do not disrupt their hosts' daily routines, are all ways to show appreciation and fulfill one's reciprocal obligation concomitantly [5,52].

Because overt and involved discussions of how guests ought to reciprocate ruin the spirit of generosity in the 'gifting' of hospitality, the terms of reciprocity tend to be murky [45]. As hosts risk providing hospitality while receiving little to no return, guests risk reciprocating in ways that go against their self-interest. As with Airbnb, sexual interactions periodically occur between hosts and guests on Couchsurfing. However, the obligatory quality of guest-host interaction and the immense power hosts hold in directing these interactions make sexual meaning-making more complicated in the context of Couchsurfing. Might guests feel pressured to reciprocate sexually due to feelings of indebtedness and a general lack of control? How does this potentially coercive relational dynamic impact guests' understandings of sexual risk and violence in the context of Couchsurfing?

More importantly, how might platform safeguards and affordances (i.e., actions made possible by platform configurations such as matching or referencing systems [18]) shape users' responses when sexual harassment or violence occurs on Airbnb or Couchsurfing? Lampinen and Cheshire's [37] research shows that hosts actively turn to Airbnb for help when problematic situations arise. Given the availability of safeguards, might Airbnb guests who have been sexually violated be equally motivated to rely on Airbnb for assistance, or do they seek other solutions? And what about Couchsurfers? Unlike Airbnb, Couchsurfing's current monetization model limits the platform's financial capacity to implement viable user protection programs. As such, how do Couchsurfing's exchange environment and lack of safeguards impact guests' experience? Exploring these questions not only provides a comprehensive look at how sexual risks are distributed and experienced within network hospitality but also lends us an empirical window into how modes of exchange and safeguards play an essential role in shaping platform outcomes and experiences in the sharing economy at large.

## 3 METHODS

To answer the questions above, I draw on interviews with 40 female Airbnb and Couchsurfing users. Like women, men also experience sexual risks in network hospitality, but Airbnb and Couchsurfing's sexual violence problem disproportionately affects the former group [12]. The decision to sample women exclusively stems from the imperative to gain a deeper understanding of their experiences to inform platform design choices that could help ameliorate existing issues for a specific target group (following a strategy similarly deployed by Furlo et al. [21]).

### 3.1 Participants & Recruitment

26 of the 40 interviews featured in this study belong to a broader study examining the nature of sexual risks in network hospitality. Between 2016 and 2021, I interviewed 96 Couchsurfers from 19 countries. The participants for this bigger study were recruited online via Couchsurfing.com and offline via in-person Couchsurfing events and gatherings in various cities across Asia, Australia, Europe, and the U.S., as permitted by public health and Institutional Review Board guidelines. Some individuals were recruited because they publicly identified themselves as victims of sexual harassment or assault in a mix of online forums, social media platforms, or news outlets. Although the sample is international, most participants were educated adults who spoke English, reflecting Couchsurfing's primary demographic [11].

These initial 26 interviews were chosen because participants invoked their opinions and experiences of sexual risks in *both* Couchsurfing and Airbnb (meaning these 26 participants were all dual-platform users). The juxtapositions these interviewees made of the platforms not only revealed critical details of how their experiences converged and diverged, but they also provided important evidence suggesting the roles that modes of platform exchange and institutional safeguards play in shaping such experiences. To dive deeper into these inductively derived perspectives, I re-interviewed 15 of these 26 individuals between 2019 and 2022 with more targeted questions on their experiences navigating sexual risks. Finally, to further triangulate my findings, I interviewed another 14 dual-platform participants recruited through network hospitality forums, Facebook groups, and snowball sampling. To summarize, the study sample consists of 55 interviews with 40 unique dual-platform users. Demographically, the overall sample leans young (mean age of 27), heterosexual (n=36), and white (n=30). See Table 1 for more details.

### 3.2 Data Collection

The semi-structured interviews were conducted in person (n=16) or on a video conferencing platform like Skype or Zoom (n=24). They lasted between 30 minutes and two and a half hours, averaging 75 minutes. When probing participants on their Airbnb experiences, I asked them expressly to reflect on experiences in shared accommodations where hosts were around to ensure fair comparison against Couchsurfing (where hosts are almost always present). To elicit descriptive statistics around safety perceptions, I also asked participants several comparative (e.g., which platform subjects you to more heightened sexual pressure?) and safety rating questions (e.g., on a scale of 1 to 5, rate how risky the search process is for both platforms?).

Table 1: Demographic details of interview participants (N=40)

| Alias | Gender identity | Sexual orientation | Ethnicity | Nationality | Age |
|---|---|---|---|---|---|
| Savannah* | Cisgender | Heterosexual | White | United States | 24 |
| Mollie | Cisgender | Heterosexual | White | United States | 20 |
| Julia* | Cisgender | Heterosexual | White | United States | 40 |
| Gillian* | Cisgender | Heterosexual | White | Canada | 26 |
| Christine* | Cisgender | Heterosexual | White | Canada | 22 |
| Mary | Transgender | Heterosexual | Hispanic | United States | 21 |
| Eva | Cisgender | Heterosexual | White | United States | 22 |
| Maggie | Cisgender | Heterosexual | White | United States | 42 |
| Mia* | Cisgender | Heterosexual | White | Italy | 22 |
| Betty | Cisgender | Heterosexual | White | United States | 20 |
| Ada* | Cisgender | Queer | White | United States | 34 |
| Jesse | Cisgender | Heterosexual | White | France | 34 |
| Ashley | Cisgender | Heterosexual | White | United States | 29 |
| Ziyi* | Cisgender | Heterosexual | Asian | Taiwan | 28 |
| Cait* | Cisgender | Heterosexual | White | Canada | 29 |
| Jackie | Cisgender | Bisexual | White | United States | 28 |
| Leena* | Cisgender | Heterosexual | White | United States | 40 |
| Hannah | Cisgender | Heterosexual | White | United States | 30 |
| Lee* | Cisgender | Heterosexual | Mixed | Singapore | 37 |
| Leila | Cisgender | Bisexual | White | United States | 24 |
| Jade* | Cisgender | Heterosexual | White | United Kingdom | 47 |
| Anna | Cisgender | Heterosexual | White | Singapore | 24 |
| Amy | Cisgender | Heterosexual | White | United Kingdom | 35 |
| Alice | Cisgender | Heterosexual | White | United States | 20 |
| Natalie* | Cisgender | Heterosexual | White | Portugal | 24 |
| Sophie* | Cisgender | Lesbian | White | United States | 36 |
| Agnes | Cisgender | Heterosexual | White | United States | 22 |
| Margaret | Cisgender | Heterosexual | White | United States | 19 |
| Edith | Cisgender | Heterosexual | Black | United States | 19 |
| Carly | Cisgender | Heterosexual | White | United States | 23 |
| Simone | Cisgender | Heterosexual | White | France | 24 |
| Francis | Cisgender | Heterosexual | White | United States | 21 |
| Ivy* | Cisgender | Heterosexual | White | United States | 26 |
| Cynthia | Cisgender | Heterosexual | White | United States | 23 |
| Caroline | Cisgender | Heterosexual | White | Canada | 20 |
| Taylor | Cisgender | Heterosexual | White | Australia | 27 |
| Paula* | Cisgender | Heterosexual | White | United States | 31 |
| June | Cisgender | Heterosexual | White | Malaysia | 29 |
| Maria | Cisgender | Heterosexual | White | Poland | 28 |
| Olivia | Cisgender | Heterosexual | White | Australia | 23 |

* Asterisk indicates participants who were interviewed twice (n=15).

Because inquiries around intimacy and sexual risk could trigger interviewees with traumatic experiences, I used various protective measures to prevent re-traumatization or stress [72]. For one, I framed my questions in a sensitive manner and was especially conscious about reminding participants of their right to pause or end the interview at any time. This measure was particularly important when conducting second-round interviews, for they often dove deeper into potentially traumatic events. Moreover, drawing on the advice of an experienced clinical researcher, I devised a distress protocol in case topics raised provoked traumatic stress reactions (see supplementary material). Although I never had to rely on this protocol, I went into each interview with a prepared list of locally relevant sexual violence support resources that I could share with my participants. Within 24 hours of every interview, I checked in with each interviewee with a thank-you email. Details that could lead to inadvertent identifications of any specific individuals have been altered to protect the identities of study participants and their hosts, and all names used in this article are pseudonyms.

All audio-recorded interviews were transcribed with the help of a research assistant. Subsequently, the research assistant and I deployed open coding to distill emergent ideas and concepts, followed by focused coding to find thematic patterns in the experiences of those interviewed using NVivo [10].[1] After the initial round of independent, focused coding, we compared levels of agreement versus disagreement in each interview and found that the inter-coder reliability was high (i.e., the ratio of the number of agreements against the sum of agreements plus disagreements was over 90%). In areas where consensus where we did not have complete consensus, we engaged in open discussions to resolve all instances of disagreements [14].

## 4 FINDINGS

To understand how Airbnb and Couchsurfing guests navigate sexual risks on the respective platforms, I break down network hospitality episodes into three phases: 1) search—finding a place to stay, 2) stay—the experience of being hosted, and 3) recourse—how post-violence experiences shape up.

### 4.1 The Configuration of Guest-Host Relations in Search

We start with the search and sort process that marks the beginning of any trip. Aside from the fact that all participants intend to minimize risk, the way search gets enacted diverges significantly based on the platform. While Airbnb users prioritize the hunt for good "properties," Couchsurfers pursue good "hosts" [33]. This schema translates into how initial guest-host relations are defined. When conducting a search, Airbnb users think of themselves as protected purchasers of a service or product, casting initial guest-host relations into a buyer-seller arrangement. Conversely, Couchsurfers think of themselves as self-responsible individuals seeking hospitality assistance. As a part of a reciprocal relationship, Couchsurfing guests see themselves as receivers and hosts as donors of a tangible benefit. Such distinctions not only underscore the pertinence of exchange and safeguards in shaping sexual risk, but they further guide platform design, affordances, and user behavior in Airbnb and Couchsurfing (see Table 2 below for a summary).

Table 2: Comparing Search on Airbnb and Couchsurfing

|  | Airbnb | Couchsurfing |
| --- | --- | --- |
| **What guests search for** | Good properties | Good hosts |
| **Role of hosts** | Logistical; service provider | Social; benefit giver |
| **Reviews** | Quantitative and qualitative | Qualitative |
| **Which platform is riskier when it comes to search?** | 7.5% (n=3/40) | 92.5% (n=37/40) |

---

[1] In this process, both the lead author and research assistant have their own access to the same interview transcripts on separate NVivo accounts. After independently completing a round of open coding on ten transcripts, we came together to discuss and align on prominent themes and categories. Then, to facilitate focused coding, we merged related codes and clearly defined them to generate the final code list. Subsequently, we used this list to code all 55 transcripts in this study consistently.

*4.1.1 Airbnb.* In the context of Airbnb, participants generally express that their interactions with hosts in shared or private room situations tend to be logistically oriented. Interview data suggests that the lack of extended social exchange makes interactions of a sexual nature rare. This finding, consistent with past research on Airbnb guest-host relations (see [33] for an example), evinces one reason why an overwhelming number of interviewees (n=37) regard Airbnb as sexually safer than Couchsurfing. Because Airbnb guests and hosts are involved in a binding negotiated exchange, where an initial payment formalizes the terms of engagement, most interviewees conduct a search assuming guest-host interactions will remain *voluntary* and *minimal*. As one participant stated, while doing extended traveling, she often pays for Airbnb rentals to break from long stretches of Couchsurfing so that she could be "left alone." Negotiated exchange, thus, reifies the distance between guest and host, dampening the chances of unwanted sexual contact.

This 'peace of mind' is further bolstered by the assumption that Airbnb, as a platform, vets hosts to ensure user safety, signaling the salience of institutional trust [2]. While Airbnb does, in fact, deploy routine AI-assisted background checks, only four participants could articulate this exact mechanism at play. The assumption that Airbnb *does something*, however, is powerful enough to motivate Airbnb guests to prioritize non-human factors such as property rating, affordability, location, amenities, and cancellation policy during the search process. Compared to Couchsurfing's search mechanism, where potential hosts' pictures and self-descriptions take center stage, Airbnb's primary search interface entirely omits information about hosts. Retrieving this information requires guests to click on individual listings and scroll to the bottom of the page. By centering product-oriented categories, Airbnb's search design reinforces the idea that the quality of the place eclipses the quality of the host [32]. Consider the opinion of Ivy (26):

> "So, for me, things like how much the place costs and how central it is matters more to me than anything. And the reviews of the place too, of course … The only time I think about the host is when I'm renting a private room in a shared space … As a woman, when I'm traveling by myself, I'll check the host's gender … Hmm, the reviews rarely say anything useful about hosts, to be frank. But I've stayed with many male hosts and have been okay, so it's not too big of a deal for me."

Many participants echoed Ivy's sentiment, noting that sexual risk assessment while searching on Airbnb only comes to mind when securing a shared or private room with a male host living on the same premises. Beyond that, they regard their search process as relatively mundane (one participant likened it to 'shopping'). Like Ivy, many guests also stressed that they deploy the additional precautionary measure of scanning listing reviews. However, they do so to gather descriptive and experiential evaluations of rentals rather than to learn more about hosts. As multiple interviewees noted, most of the reviews on Airbnb lean positive [70] and rarely contain "meaningful information" about abnormal hosting behavior. Simone (24) suggested that part of the reason why this might be the case has to do with the safeguards Airbnb offers:

> "I don't think people bother leaving bad reviews on hosts because if they try something funny with you, you will probably try and remove yourself from the situation and let Airbnb know? And they'll likely give you a refund or credit to stay elsewhere. Not sure if there is an incentive to leave a negative review after that, so reading reviews for sexual safety feels moot."

*4.1.2 Couchsurfing.* Search on Couchsurfing could not be more different. Lodged in an alternative exchange system and without formal safeguards, what lies between Couchsurfing guests and hosts are reciprocal exchanges based on individual goodwill and the principles of gifting [25]. Insofar as allowing numerical ratings for hosts and living spaces risks corrupting the sacred nature of gifted hospitality, Couchsurfing does not allow guests to rank hosts' profiles or accommodations based on any 'objective' quality measures. Moreover, contrary to Airbnb, references on Couchsurfing contain primarily qualitative information [42], with reviewers focusing on the quality of *guest-host interactions* instead of the accommodation itself. Extending this insight to searching for a host, Couchsurfing guests invest significant time and cognitive energy studying profiles and references to ascertain if their host might be a good match. In other words, the heightened risk involved in sustained social interactions between givers (i.e., hosts) and receivers (i.e., guests)—two parties with significant power differentials—compels guests to believe *who* they stay with matters significantly more than the place itself. Exemplifying this idea, one participant said that she would rather "stay with a kind host in the suburbs than a dodgy one in a fancy area."

Because of the risks involved in sharing space and engaging in reciprocal exchange with strangers, many participants stress that gender pairing is something they seek to control to mitigate sexual risks. For example, Ada (34) and Ziyi (28) offered two explanations highlighting why gender matters in the search phase:

> *Ada*: "This is literally the most important thing to think about when I look for a host. If a guy only hosts women, it's a major red flag for me. Like, what do you want from me? That's an immediate red flag. I avoid them altogether. I'd only send requests to men if he hosts both men and females. Sometimes some hosts say on their profiles that if you stay with them, you will have to share their bed. That's also a major red flag … Most of the time, I exclusively request to stay with female hosts if I'm by myself."

> *Ziyi*: Sometimes, on male profiles, they list themselves as a nudist or someone who enjoys massage exchange. This is becoming more and more common. Many men want you to be naked with them at home. I don't need to tell you why they like that. I mean, it's fine if that's your thing, but even if that's not, you're probably gonna feel the pressure to do it. It's their place and they are letting you stay for free, so it's very difficult to get yourself out of these situations. If you are too reserved or shy and it pisses them off, they could kick you out … It sucks because it's not like Couchsurfing will come to your aid."

By asking, "what do you want from me," Ada suggests that male hosts who exclusively host women must have an ulterior motive, dismantling the disinterested notion of hospitality gifting and revealing the potentially manipulative and coercive nature of exchange in Couchsurfing. This question not only accentuates the workings of reciprocal exchange in shaping how Couchsurfers search for hosts, but it also shows how inhabiting a social space where exchange terms are ambiguous, staying with certain types of hosts may lead to heightened sexual risks. Ziyi, who described the "pressure" one might feel to be naked when staying with nudists, cogently highlights the sexual precarity Couchsurfing guests find themselves in—that in an arrangement where they get to "stay for free," they may be compelled to do things against their will to appease their hosts. Because safeguards are absent in Couchsurfing, upsetting one's host could lead to temporary homelessness for guests, a costly consequence for those traveling on a budget. For these very reasons, two respondents in this study mentioned that after using Couchsurfing for several years, they now almost exclusively search for female hosts to avoid issues with men.

When asked to compare the search process between Airbnb and Couchsurfing, many participants mentioned that the cost of picking the wrong host on Couchsurfing is definitively heftier than choosing the wrong one on Airbnb. Jade (47), a seasoned Couchsurfer, explained that it is "crucial to pick good hosts" in Couchsurfing because their presence could "literally make or break your trip." Echoing her sentiment, another interviewee said that "mistakes made in Couchsurfing are more consequential since it's not Airbnb … if a weird sex thing happens, you don't just get a credit to stay elsewhere." Not having binding negotiated exchanges and safeguards not only puts network hospitality users in a riskier situation, but it also makes them think that the onus of ensuring one's sexual safety is entirely on themselves.

## 4.2 In-person Sexual Scripts and Meaning-making

Modes of exchange and safeguards play an equally important role in shaping how in-person sexual encounters between hosts and guests transpire. In the context of Airbnb, guests who have had sexual encounters with their hosts generally describe their experiences in two distinct ways—that it is either a clear case of consensual intimacy or sexual violence. The former strengthens guest-host relationships, while the latter calls for an immediate activation of platform safeguard protection. The sexual scripts, alongside how users make meaning of these interactions, are stabilized by the binding negotiated exchange Airbnb facilitates. On the other hand, even though consensual sex also exists in Couchsurfing, sexual scripts and meaning-making are a lot more contingent in this context, ultimately increasing relational ambiguity and the overall sexual risk guests assume (see Table 3 below for a summary).

Table 3: Comparing In-person Sexual Scripts and Meaning-making on Airbnb and Couchsurfing

|  | **Airbnb** | **Couchsurfing** |
| --- | --- | --- |
| **On a scale of 1-5, how obligated do you feel to hang out with your host?** | 0.9 (mean) | 4.3 (mean) |
| **On a scale of 1-5, how standardized is the interactional script on each platform?** | 4.1 (mean) | 2.5 (mean) |
| **Which platform subjects you to more heightened sexual pressure?** | 2.5% (n=1/40) | 97.5% (n=39/40) |
| **Which platform is riskier when it comes to in-person interactions?** | 5% (n=2/40) | 95% (n=38/40) |

*4.2.1 Airbnb.* While initial interactions between Airbnb guests and hosts may resemble buyer-seller relations, sustained engagement between two parties may lead to varying degrees of physical closeness. More specifically, the voluntary nature of Airbnb sociality encourages users to, as one interviewee puts it, "hang out without any agenda." As Leena (40) described, the development of intimacy between two consensual adults in such circumstances is anything but extraordinary:

> "There is, of course, a chance that two people might develop chemistry, flirt a little here and there, and form a connection. Sometimes, something happens right there and then. But the intimate stuff is usually desired by both parties … I feel like, in general, I'm able to block advances I don't appreciate and pursue whatever only if I really wanted to. So, all in all, I do tend to feel quite safe … Reason is simple, *I paid for the* *room*, and I don't think any sensible host would want to get blacklisted by Airbnb for doing something stupid. So yeah, I have a bit more control *over the situation*." (emphasis mine)

Many participants echoed Leena's assertion that binding negotiated exchange gives guests more power to decide how much socializing they want to engage with their hosts. As such, if a guest finds genuine interest in interacting with a host and decides to do so, transformative and meaningful connections—both sexual and non-sexual in nature—frequently arise. This finding parallels those found by Ikkala and Lampinen [29], who similarly suggest that negotiated exchange in network hospitality clears a path for rewarding sociality by mitigating one's sense of obligation. Unexpected sexual connections not only add value to one's trip, but they also remind participants, as one of them puts it, of the "human side" of a platform that in recent years has become increasingly "standardized and hotel-like" (see [48,56] for similar critiques).

However, on rare occasions, seemingly non-obligatory guest-host interactions might motivate non-consensual sexual encounters. Many women in this study mention that the voluntary nature of their interactions with male hosts sometimes gives men the idea that they are interested in more than platonic interactions. In other words, some hosts may harbor the idea that guests who socialize with them do so because they are either romantically or sexually attracted to them. Paula (31), for instance, disclosed that she has had several encounters with hosts who misconstrued a friendly hang-out as a casual date: "They can get so physical and flirtatious, and sometimes it's really uncomfortable." In many of these risky situations, navigating one's way out can prove emotionally enervating. However, most participants who had unwanted sexual advances directed at them mentioned that they had successfully rebuffed overly flirtatious hosts by subtly communicating their discomfort and closing the door to their rooms to limit further contact. "It feels a bit awkward at times to disengage like that," said one participant, later adding, "but I have to remember that I don't owe them (hosts) anything." She then elaborated that "shutting the door on one's host" is a "luxury" one simply does not have in Couchsurfing.

*4.2.2 Couchsurfing.* Reciprocal exchange in Couchsurfing makes sexual scripts and meanings attributed to physical encounters less structured and more variegated. For one, without formalized or contracted expectations around how to organize guest-host exchanges, Couchsurfing guests sometimes find it hard to articulate if certain behaviors are born out of their autonomous will or a compulsion to perform appreciation for receiving hospitality. Placing sex within this exchange system makes for a variety of meanings and emotional outcomes that arguably turn Couchsurfing into a riskier sexual terrain for guests to navigate.

To begin, it is important to note that many Couchsurfers have consensual sex with their hosts, not unlike what Leena described above vis-à-vis Airbnb. When recounting these encounters, Couchsurfers allude to the intimate settings guests and hosts share and how getting to know each other could induce sexual chemistry that leads to physical exchange. However, when invoking such experiences, Couchsurfers are cognizant of the potentially problematic interpretations others might have of these events. For instance, Maggie (42), a global trotter who has hooked up with more than 20 hosts, said that even though she conceives of sex between guests and hosts as "normal," she is mindful that some people might construe these sexual occurrences as "a form of prostitution." This perception, she said, is a problem unique to Couchsurfing and "does not apply to Airbnb" because the latter includes a financial transaction that precludes a morally questionable form of sexual reciprocity. By invoking what Zelizer calls the "hostile worlds" perspective [69], Maggie underscores a general morality that pegs the exchange of economics and intimacy as taboo. Amy (35), another interviewee who raised a similar concern about sexual contact between guests and hosts, added that at times, she questions if she might hook up with some of her hosts if she had met them in another setting:

> "I don't want to say I was ever forced, but there is definitely some invisible pressure at times. For two or three of these hosts, if I had met them at a bar, I'm not sure if I'd go home with them … I can't read the minds of these men, so I don't know if they think maybe part of why we hooked up is due to the whole Couchsurfing setup. But if you ask them, they'd be lying if they'd told you that the dynamics don't matter. This is maybe how some men get to hook up with hot girls ha!" (emphasis hers)

An underlying tit-for-tat mentality that undergirds reciprocal exchange in Couchsurfing—which one participant suggests "nobody talks about, but everyone knows"—seems to make the notion of voluntary behavior untenable for many participants, ultimately impacting how sexual meanings are constructed. For instance, Lee (37), a veteran Couchsurfer with eight years of experience, mentioned that during her first experience as a guest, she rebuffed her male host's demand to sleep on the same bed the night they had met. Her host appeared visibly irritated during breakfast the next day, making their subsequent interactions awkward and uncomfortable. That night, while watching television on the couch, Lee did not resist her host's attempt to kiss and cuddle up with her. "I had another day with him and basically allowed it to ease things up," she said. Reflecting on her experience, Lee found it difficult to interpret what had transpired on the second night, saying that while "it did not feel non-consensual," it also did not make her "feel good."

Like Lee, many other women in the study emphasize the complex and ambiguous nature of sexual exchange on Couchsurfing when describing their own encounters. In many of these accounts, participants stress that as guests, they feel obliged to socialize with and appease the needs of hosts and that while most hosts are accommodating and respectful, others are less so. In cases where unpleasant sexual encounters transpire between participants and their hosts, guests are slow to jump to the conclusion that the occurrences were non-consensual or to be classified as sexual violence. When probed, many women suggest that the logic underpinning Couchsurfing's exchange system and the general lack of safeguards play a role in fostering this sense of ambiguity. As one puts it, it can be "challenging to say no to someone who is the only reason you have a shelter that night." Echoing this opinion is Gillian (26), an ex-Couchsurfer who recently left the platform because of a sexual violence incident. Assaulted by her host hours after they met, Gillian stayed the night after the incident because she feared the consequences if she had left. In the following months, Gillian struggled to define what had happened to her, saying that she probably would have left immediately if that had been a clear case of sexual assault. However, after close to half a year of therapy, she recently concluded that the "inherently coercive" exchange context that hosts and guests traverse, especially if it involves a man and a woman, renders most sexual encounters in Couchsurfing non-consensual:

> "When I think back to that night, it's either do what he wants or get kicked out, and I'll be wandering on the streets in the middle of the night. Either way, I'm fucked … Consent is so fraught when you are put in a situation where you have no options."

Comparing sexual risks between Airbnb and Couchsurfing, 38 out of 40 participants believed that the unpredictability of social interactions brought about by the dynamics of reciprocal exchange on Couchsurfing produces a riskier sexual environment for its users. In addition, although the nature of sex in Couchsurfing, akin to Airbnb, contains both clear-cut cases of consensual intimacy and sexual violation, many more incidents are interpreted as meaningfully and consensually ambiguous. This dynamic not only makes Couchsurfing less safe, but the stark psychological and physical costs many individuals pay while using the platform make the idea of 'free' hospitality in Couchsurfing tenuous.

### 4.3  Post-violence Reactions and Recourse

In the event of sexual violence, Airbnb and Couchsurfing users chart distinct strategies concerning who they turn to and how they seek justice. Here, modes of exchange and platform safeguards continue to play a crucial role. To start, even though sexual violence occurs more frequently on Couchsurfing, sexual violence is more frequently reported to Airbnb than on Couchsurfing (75% vs. 11.5%, respectively—see Table 4 for a summary). While a key motivation for users to make such reports to Airbnb is to encourage the platform to sanction perpetrators and delist their properties, many survivors do so also to access the benefits associated with safeguard protections, which include monetary compensation and rebookings. In other words, Airbnb's safeguards provide structure and formalize sexual violence recourse for users. As one participant puts it, "It's what you do when you are not satisfied with something you've bought, you contact customer support." Couchsurfers, on the other hand, are less likely to report sexual violence formally to the platform and are comparably more haphazard in their post-incident strategizing. Recognizing that they pay little to nothing to access Couchsurfing and that no protection schemes are put in place, Couchsurfers do not expect the platform to intervene in a substantial manner and rely on other means to help themselves get to a safe place and acquire a sense of justice.

Table 4: Comparing Post-Violence Reactions and Recourse on Airbnb and Couchsurfing*

|  | Airbnb | Couchsurfing |
| --- | --- | --- |
| **Reporting to the platform** | 75% (n=3/4) | 11.5% (n=3/26) |
| **Financial compensation after reporting to the platform** | 100% (n=3/3) | 0% (n=0/3) |
| **Finding a new place** | Booking on another Airbnb listing or hotel (paid by Airbnb) | Couchsurfing's emergency forums to identify new hosts or hotel (paid by guest) |
| **Reporting to police** | 25% (n=1/4) | 4.5% (n=1/26) |

* This table only applies to those who self-reported having experienced sexual violence on either Airbnb (n=4), Couchsurfing (n=26), or both (n=3). Note that self-evaluations of sexual violence can be a highly subjective and culturally dependent process [26], and even participants who did not express having experienced sexual violence discussed their dealings with sexual risks on these platforms at length.

*4.3.1 Airbnb.* Three out of four Airbnb guests in this study who have had first-hand experience with sexual assault recalled contacting Airbnb directly after an incident. Rather than calling the police, many guests—particularly if they were abroad—find a way to leave their hosts' place and contact Airbnb's 24/7 customer support system instead. Most cite language barriers, a lack of knowledge of local laws, and the hassle of navigating a criminal case abroad as critical reasons for sidestepping the police. Three other factors further propel women to turn to Airbnb instead: participants want the platform to 1) assist in relocating them to a safe place in the shortest possible time, 2) sanction the perpetrator, and 3) provide financial compensation or legal advice. Most of the time, participants report that Airbnb refunds their payment and provides instant credit for booking an accommodation at another Airbnb property or hotel. Although few participants attribute fault to Airbnb for what had occurred to them, most believe that Airbnb remains responsible for what happens to guests after they have been assaulted, underscoring the instrumental function of an exchange system that guarantees safeguards.

While platform safeguards allow users to find refuge immediately, they could dampen an individual's will to pursue further actions. As one sexual assault survivor said:

> "When I mentioned what had happened and asked if I should contact the police, the Airbnb rep avoided my question and said she could help me find a place that night and that Airbnb would comp the stay. It felt like they were basically throwing money at me so I wouldn't blow it up. They wanted me to just stop there, and I guess I kinda did."

Another survivor, Cait (29), later discovered that her host was a registered sex offender weeks after she had launched a formal complaint to Airbnb. Upset that the platform had betrayed her trust, Cait was perplexed by how "a company so profitable" was not "able to sort out such a major loophole" in their screening process. Ultimately, Cait's dispute with Airbnb ended with a settlement she deemed reasonable. We could glean from these cases that while binding negotiated exchange and safeguards provide those who suffer the brunt of sexual assault with immediate recourse, they curtail the pursuit of further actions that could lead to greater justice. By enabling the company, as an ex-employee told *Bloomberg Businessweek*, to "shoot "the money cannon" at problems" [9], Airbnb minimizes legal troubles and reputational costs at the expense of implementing more extensive programs that lead to systemic protection. While occasional loopholes fly in the face of Airbnb's self-proclaimed safety prowess, most participants of this study remain appreciative of the structured and predictable nature of the platform's aid: "It's not perfect and nothing is, but you can tell that they have your interest in mind" (Taylor, 27).

*4.3.2 Couchsurfing.* When Couchsurfers encounter sexual violence during their stay, they do not turn to the platform like Airbnb users. Jesse (34) explained why:

> "I don't contact them because I don't think they can do anything for me. Do they even have a number I can call? I mean, the company is broke, nobody's paying, so I highly doubt they even have a real customer service team … The last time I checked their safety page, I recall a link to an email. Who is going to write them an email when you are in dire danger? Therefore, I don't bother."

Like Jesse, many Couchsurfers hint at the futility of reporting sexual crimes to Couchsurfing due to its perceived lack of customer support infrastructure and resources. Without the safeguards accompanying binding negotiated exchange, Couchsurfers recognize the limited extent to which their platform could intervene and provide economic support or recourse to those with urgent needs. This perspective is equally shared by those who have paid a $60 lifetime verification fee or the recently implemented monthly or yearly membership fees. More specifically, these users conceive their monetary contribution to Couchsurfing as a donation to keep the platform alive rather than a payment that accompanies user protection. Some Couchsurfers earmark money paid to Couchsurfing as an *access* fee, distinct from service fees Airbnb charges with each transaction (which most regard as a form of financial assurance that grants users protection for each stay).

Without platform safeguards to rely on, Couchsurfers who experience sexual violence respond in various ways. For one, those with more financial flexibility find tactful ways to leave and seek accommodation at hostels, hotels, or Airbnb properties. Conversely, those traveling with tighter budgets may either send last-minute requests to hosts in the area or turn to emergency discussion groups on Couchsurfing's website to locate hosts willing to take them in. An interviewee theorized that the ubiquity of sexually exploitative hosts in part explains why these emergency discussion groups have become so popular over the years. On rare occasions, some have even abandoned the rest of their vacation plans to return home.

As with Airbnb users, Couchsurfers rarely contact the police after experiencing sexual violence except for the most severe of cases (through self-definition). This pattern of behavior echoes previous research on sexual assault, which suggests that survivors may not come forward for long periods due to shame, self-blame, or a lack of recognition that what they had undergone classifies as rape [63,64]. The tendency to blame oneself for picking a riskier way to travel, combined with a lack of clarity around how consent ought to be interpreted, contributes to the abovementioned patterns. Moreover, lacking access to timely financial compensation, combined with the desire to quell further emotional fluctuations during traveling, survivors in this study generally wait until the conclusion of their trip to decide what actions to take.

Although a small number of Couchsurfers eventually report these occurrences to Couchsurfing (11.5%—3 out of 26 survivors in this study), which has occasionally led to Couchsurfing removing the perpetrators' profiles, many more Couchsurfers pursue their own version of justice by leaving negative references on their hosts' profiles. On a platform where references are overwhelmingly positive [7], a single negative reference could severely jeopardize one's surfing and hosting prospects [42]. Furthermore, because survivors understand that in a reciprocity-based community where host selection and perusing references have heightened importance, the costly nature of writing negative references is seen as a more direct and meaningful weapon one could wield to punish their perpetrators. Below, Julia (40) explained her motivation behind why she left her abuser a negative reference:

> "I wanted everyone to know what he did to me. I know many people avoid negative references because their hosts could retaliate by doing the same, but I cannot stand the thought of him abusing other women. It just sickens me … I noticed that after I left my reference, he deleted his profile and made a new one, so I wrote the same reference. It is now my life's mission to destroy his chances of hurting other surfers again."

## 5 DISCUSSION & CONCLUSION

Drawing on interviews with 40 female Airbnb and Couchsurfers users, this article addresses the complex relationship between sexual risk, modes of exchange, and safeguards in two prominent network hospitality platforms across three stages—search, in-person interactions, and post-violence reactions. I argue that Airbnb's provision of binding negotiated exchange and institutional safeguards reduces sexual risk by casting initial guest-host relations into a buyer-seller arrangement, stabilizing interactional scripts, and formalizing sexual violence recourse. Conversely, Couchsurfing's facilitation of reciprocal exchange, alongside the lack of safeguards, increases sexual precarity both on- and off-platform. Couchsurfers overwhelmingly express the sentiment that the responsibility of sexual risk mitigation lies squarely in their own hands and that it is up to them to navigate their way out of unpleasant sexual situations. These dynamics not only manifest through platform design and affordances, but they also shape how users interface with their communities both digitally and offline.

### 5.1 Sexual Risk, Exchange, and Safeguards on Platforms

By priming users on what to prioritize during a search, how to engage in social exchange with hosts offline, and what to do when they fall victim to sexual violence, exchange and safeguards shape sexual risk negotiation in network hospitality across multiple stages. When it comes to search, Airbnb users focus more on the property than host characteristics [33], trusting that the platform has done its due diligence in vetting hosts. Akin to Lampinen and Cheshire [37], who demonstrate the counterintuitive way binding negotiated exchange and safeguards contribute to prosocial interactions, my research illustrates that the voluntary nature of hosts and guests' in-person interactions indeed creates space for positive and consensual sexual exchanges to occur. In the relatively unlikely event of sexual violence, users actively turn to Airbnb for both financial and logistical support.

Sans negotiation and financial assurances, reciprocal exchange between Couchsurfing hosts and guests produces more contingencies and risks for users. During the search phase, users spend extensive time and energy reading profiles and identifying viable hosts. While reciprocal exchange catalyzes prosocial forms of sexual exchange, it concomitantly increases the sexual risk of guests who feel a general sense of indebtedness. As Sahlins [54:207] reminds us, reciprocity facilitates the "formation of rank itself," whereby power is bestowed on those who are generous. In a space where reciprocal pressures are strong and the power imbalance between hosts and guests is stark, guests often find it hard to articulate if sex is born out of one's autonomy or compelled through Couchsurfing's relational setting. As described by many participants of this study, the opacity of consent in Couchsurfing points to the challenges of making quick decisions and precise interpretations when unpleasant sexual encounters occur. In the end, without appropriate platform safeguards, the heightened perception of risk increases one's sense of self-responsibility, propelling Couchsurfers to think that the onus of ensuring one's sexual safety falls entirely on one's own shoulders. These findings demonstrate the inextricable connection between platform exchange, safeguards, and sexual meaning-making.

For conceptual clarity, I would like to underscore the link between modes of exchange and the presence of safeguards in platforms. In some digital spaces, multiple modes of exchange exist. For instance, through websites like Craigslist, one could either pay or trade to acquire something. However, because Craigslist neither binds negotiated transactions nor offers safeguards, disputes—if and when they arise—typically stay between transactants. In this context, exchange partners generally assume all risks involved in a transaction. To lower such risks, users may have to pay more for the services of a binding platform. While such platforms take a financial cut from these transactions, they also use the payments received to fund user protection programs. Such systems reduce the perception of risks for users and codify a deeper sense of institutional trust. By charging a slight premium on business transactions, third-party platforms such as Venmo or PayPal rely on this very mechanism to mitigate exchange risks for their users. As I have demonstrated throughout this study, binding negotiated exchange and the safeguards it affords provide users with deep psychological comfort, ultimately reducing interactional and cognitive stress associated with risk mitigation and management.

The findings of this study also suggest that while studying sexual risk in network hospitality, we need to go beyond the moral distinction between commercial versus noncommercial forms of exchange and pay attention to other underlying mechanisms, such as the risks involved in whether the terms of exchange have been bound by a third-party and if safeguards were present [13,44]. For instance, I find that Couchsurfing, a platform many deem to be the poster child of the sharing economy for its reliance on non-monetary forms of exchange, fosters exploitative behaviors between users in a way more systemic than what we witness on Airbnb. Although conventional economic wisdom suggests that reciprocity typically leads to more prosocial outcomes, the excessively social nature of Couchsurfing can engender interactional risks that ultimately jeopardize, instead of strengthen, sociality.

## 5.2 Designing for Vulnerable User Populations

Akin to prior research [49,60], this study calls further attention to addressing the needs of vulnerable populations in network hospitality and the sharing economy at large. By highlighting how platform organization and design structurally increase risks for certain populations while exacerbating existing forms of inequalities experienced in a broader societal context, I argue that critiquing technology's role in shaping interpersonal relations during an era of platform capitalism is imperative.

Armed with the knowledge that contextual, relational, and exchange conditions work hand in hand to accentuate risks for certain populations, HCI scholars have the continued responsibility to recommend platform changes to blunt antisocial interactions. In my case, many participants welcome the idea of platforms implementing evidence-backed safety advice to help users mitigate sexual risks. More specifically, these participants assert that instead of perpetuating an "asexual imperative" [65:33] and downplaying network hospitality's sexual violence issue, platforms are responsible for educating their users on relevant risks. For example, while Couchsurfing's current "Conduct Policy" includes a "Don't Look for a Date" clause, and Airbnb's "Safety" page consists of a line reminding users not to "commit sexual assault, sexual abuse, sexual harassment, [and] domestic violence," many participants who have seen these pages describe the information listed as sparse and superficial. What many deemed to be potentially more helpful is a resource page featuring common scenarios and a list of risk-mitigation strategies applicable to different stages of a network hospitality experience. Such a page could be assigned as 'mandatory reading' and made available during user onboarding for maximum visibility and reach.

Relatedly, platforms like Couchsurfing could also implement nudges [4,40] to warn guests of hosts with skewed hosting gender ratios. Namely, before female guests message prospective male hosts who exclusively host women, an automatic alert detailing non-paternalistic safety measures could be triggered to help them make a more informed decision. In similar contexts where users exchange hospitality through informal means and safeguards are absent, a more streamlined emergency feature that matches those who fall victim to sexual violence to verified local hosts was thought to be a potential game-changer for many. While many of these measures direct more attention to those at risk of sexual violence rather than those who perpetrate it, it is important to remember that a multi-faceted approach is needed to boost platform safety for all users. More specifically, developing mechanisms that ensure sexual predators are appropriately identified and barred in a timely fashion should go hand in hand with the recommendations above.

Additionally, from a cultural angle, designing such measures requires platform designers to acknowledge the needs of and consult with vulnerable user populations, which naturally helps elevate the perspectives of those who have traditionally been sidelined. In alignment with Zytko et al. [72], I encourage platform designers and researchers to follow the principles of feminist HCI [3]; only by involving end-users in the design process can we build more robust systems that are low in risk and protect one's sexual autonomy.

## 5.3 Limitations and Future Research

I end with a few limitations in this study, hopefully serving as a launchpad motivating future inquiries. First, given the study's qualitative focus, claims about the prevalence of consensual versus non-consensual sexual encounters in network hospitality go beyond the scope of this article. Acquiring this information would be generative for future studies and allow researchers to identify the extent of sexual risk within this economy. Holding concrete statistics on the groups that are particularly vulnerable in any given context could also help us predict sexual assault occurrence, allowing us to implement more targeted computational or human interventions to help improve overall platform safety. Because my current sample skews white and heterosexual, future work should seek to investigate the experiences of more diverse groups of participants. For example, many straight female Couchsurfers choose to stay with other women to reduce sexual risk, but how may this dynamic alter the experiences of lesbian or bisexual women?

Finally, interest in gifting, particularly within the digital realm, is growing within the field of HCI [34,58,66]. As platforms that promote alternative modes of economic participation continue to make inroads into people's lives, I hope that more attention is given to examining how gifts today produce relational intricacies in hybrid settings where people's interactions cut across digital and physical worlds. Because gifting can appear more socially oriented due to its apparent disinterestedness [41], I encourage future research on the topic to pursue more comparative interrogations. As is the case with this study, juxtaposing how reciprocity works alongside more traditional forms of economic engagement (i.e., paying) in different networked spaces invites complex discussions around what being 'social' really entails, as well as the risks and trade-offs people assume in these situations. If what Fourcade and Kluttz [20] argue holds water, that gifting and reciprocity form the cornerstone of today's digital capitalism, there is no better time to unwrap the impact of such processes on society than now.

## ACKNOWLEDGMENTS

This study would not have been possible without the research participants who so generously shared their stories and hospitality with me, as well as the funding support from the Social Science and Humanities Research Council of Canada, the Institute of International Studies, and the Social Sciences Research Pathways program at UC Berkeley. I am indebted to Marion Fourcade, Ann Swidler, Eliza Brown, Cal Morrill, Morgan Ames, Luis Tenorio, Margaret Eby, Jack Ching, members of the UC Berkeley Gender and Sexualities Workshop, and the anonymous CSCW reviewers for their thoughtful questions and suggestions. Finally, I thank Danielle Leard, Jacqueline Salguera, and Inaara Charolia for their indispensable research assistance.

## REFERENCES

[1] Ashwaq Alsoubai, Jihye Song, Afsaneh Razi, Nurun Naher, Munmun De Choudhury, and Pamela J. Wisniewski. 2022. From "Friends with Benefits" to "Sextortion:" A Nuanced Investigation of Adolescents' Online Sexual Risk Experiences. *Proc. ACM Human-Computer Interact.* 6, CSCW2 (November 2022), 1–32. DOI: https://doi.org/10.1145/3555136

[2] Reinhard Bachmann and Andrew C. Inkpen. 2011. Understanding institutional-based trust building processes in inter-organizational relationships. *Organ. Stud.* 32, 2 (March 2011), 281–301. DOI: https://doi.org/10.1177/0170840610397477

[3] Shaowen Bardzell. 2010. Feminist HCI: Taking stock and outlining an agenda for design. In *Conference on Human Factors in Computing Systems - Proceedings*, 1301–1310. DOI: https://doi.org/10.1145/1753326.1753521

[4] Kristoffer Bergram, Marija Djokovic, Valéry Bezençon, and Adrian Holzer. 2022. The Digital Landscape of Nudging: A Systematic Literature Review of Empirical Research on Digital Nudges. In *Conference on Human*

*Factors in Computing Systems - Proceedings*, ACM, New York, NY, USA, 16. DOI: https://doi.org/10.1145/3491102.3517638

[5] Paula Bialski. 2012. Technologies of hospitality: How planned encounters develop between strangers. *Hosp. Soc.* 1, 3 (2012), 245–260. DOI: https://doi.org/10.1386/hosp.1.3.245_1

[6] Lindsay Blackwell, Nicole Ellison, Natasha Elliott-Deflo, and Raz Schwartz. 2019. Harassment in social virtual reality: Challenges for platform governance. *Proceedings of the ACM on Human-Computer Interaction 3*. DOI: https://doi.org/10.1145/3359202

[7] Rachel Botsman and Roo Rogers. 2010. *What's Mine is Yours: The Rise of Collaborative Consumption*. HarperBusiness, London.

[8] Johanna Brewer, Joseph Jofish Kaye, Amanda Williams, and Susan Wyche. 2006. Sexual Interactions: Why we should talk about sex in HCI. In *Conference on Human Factors in Computing Systems - Proceedings*, 1695–1698. DOI: https://doi.org/10.1145/1125451.1125765

[9] Olivia Carville. 2021. Inside Airbnb's 'Black Box' Safety Team: Company Spends Millions on Payouts - Bloomberg. *Bloomberg Businessweek*. Retrieved September 14, 2022 from https://www.bloomberg.com/news/features/2021-06-15/airbnb-spends-millions-making-nightmares-at-live-anywhere-rentals-go-away

[10] Kathy Charmaz. 2006. *Constructing grounded theory: A practical guide through qualitative analysis*. SAGE Publications, Thousand Oaks.

[11] De-Jung Chen. 2012. Global concept, local practice: Taiwanese experience of CouchSurfing. *Hosp. Soc.* 1, 3 (2012), 279–297. DOI: https://doi.org/10.1386/hosp.1.3.279_1

[12] De-Jung Chen. 2017. Gendered Couchsurfing: women from Western Europe and East-Asia contesting de-sexualised cosmopolitanism. *Gender, Place Cult.* 24, 8 (2017), 1090–1106. DOI: https://doi.org/10.1080/0966369X.2017.1372375

[13] Coye Cheshire, Alexandra Gerbasi, and Karen S. Cook. 2010. Trust and transitions in modes of exchange. *Soc. Psychol. Q.* 73, 2 (2010), 176–195. DOI: https://doi.org/10.1177/0190272509359615

[14] Bonnie Chinh, Abbas Ganji, Himanshu Zade, and Cecilia Aragon. 2019. Ways of qalitative coding: A case study of four strategies for resolving disagreements. In *Extended Abstracts of the 2019 CHI Conference on Human Factors in Computing Systems*, Association for Computing Machinery, 1–6. DOI: https://doi.org/10.1145/3290607.3312879

[15] Ioanna Constantiou, Attila Marton, and Virpi Kristiina Tuunainen. 2017. Four models of sharing economy platforms. *MIS Q. Exec.* 16, 4 (2017), 236–251.

[16] Russell Cropanzano, Erica L. Anthony, Shanna R. Daniels, and Alison V. Hall. 2017. Social exchange theory: A critical review with theoretical remedies. *Academy of Management Annals 11*, 479–516. DOI: https://doi.org/10.5465/annals.2015.0099

[17] Prema Dev, Jessica Medina, Zainab Agha, Munmun De Choudhury, Afsaneh Razi, and Pamela J. Wisniewski. 2022. From Ignoring Strangers' Solicitations to Mutual Sexting with Friends: Understanding Youth's Online Sexual Risks in Instagram Private Conversations. In *Proceedings of the ACM Conference on Computer Supported Cooperative Work, CSCW*, Association for Computing Machinery, 94–97. DOI: https://doi.org/10.1145/3500868.3559469

[18] Jennifer Earl and Katrina Kimport. 2011. *Digitally enabled social change: Activism in the internet age*. MIT Press, Cambridge.

[19] Eyal Ert and Aliza Fleischer. 2019. The evolution of trust in Airbnb: A case of home rental. *Ann. Tour. Res.* 75, (March 2019), 279–287. DOI: https://doi.org/10.1016/j.annals.2019.01.004

[20] Marion Fourcade and Daniel N. Kluttz. 2020. A Maussian bargain: Accumulation by gift in the digital economy. *Big Data Soc.* 7, 1 (2020). DOI: https://doi.org/10.1177/2053951719897092

[21] Nicholas Furlo, Jacob Gleason, Karen Feun, and Douglas Zytko. 2021. Rethinking Dating Apps as Sexual Consent Apps: A New Use Case for AI-Mediated Communication. In *Proceedings of the ACM Conference on Computer Supported Cooperative Work, CSCW*, Association for Computing Machinery, 53–56. DOI: https://doi.org/10.1145/3462204.3481770

[22] Jennie Germann Molz. 2013. Social networking technologies and the moral economy of alternative tourism: The case of couchsurfing.org. *Ann. Tour. Res.* 43, (2013), 210–230. DOI: https://doi.org/10.1016/j.annals.2013.08.001

[23] Jennie Germann Molz. 2014. Toward a network hospitality. *First Monday* 19, 3 (March 2014). DOI: https://doi.org/10.5210/fm.v19i3.4824

[24] Juho Hamari, Mimmi Sjöklint, and Antti Ukkonen. 2016. The sharing economy: Why people participate in collaborative consumption. *J. Assoc. Inf. Sci. Technol.* 67, 9 (September 2016), 2047–2059. DOI: https://doi.org/10.1002/asi.23552

[25] John Harvey, Andrew Smith, and David Golightly. 2017. Giving and sharing in the computer-mediated economy. *J. Consum. Behav.* 16, 4 (July 2017), 363–371. DOI: https://doi.org/10.1002/cb.1499

[26] Heather R. Hlavka. 2014. Normalizing Sexual Violence: Young Women Account for Harassment and Abuse. *Gend. Soc.* 28, 3 (February 2014), 337–358. DOI: https://doi.org/10.1177/0891243214526468

[27] Ian W. Holloway, Shannon Dunlap, Homero E. del Pino, Keith Hermanstyne, Craig Pulsipher, and Raphael J. Landovitz. 2014. Online Social Networking, Sexual Risk and Protective Behaviors: Considerations for Clinicians and Researchers. *Current Addiction Reports 1*, 220–228. DOI: https://doi.org/10.1007/s40429-014-0029-4

[28] Noor Zainab Hussain and Joshua Franklin. 2020. Airbnb valuation surges past $100 billion in biggest U.S. IPO of 2020 | Reuters. *Reuters*. Retrieved September 14, 2022 from https://www.reuters.com/article/airbnb-ipo/airbnb-valuation-surges-past-100-billion-in-biggest-u-s-ipo-of-2020-idUSKBN28K261/

[29] Tapio Ikkala and Airi Lampinen. 2015. Monetizing network hospitality: Hospitality and sociability in the context of Airbnb. In *CSCW 2015 - Proceedings of the 2015 ACM International Conference on Computer-Supported Cooperative Work and Social Computing*, 1033–1044. DOI: https://doi.org/10.1145/2675133.2675274

[30] Shagun Jhaver, Yoni Karpfen, and Judd Antin. 2018. Algorithmic anxiety and coping strategies of airbnb hosts. In *Conference on Human Factors in Computing Systems - Proceedings*. DOI: https://doi.org/10.1145/3173574.3173995

[31] Jiwon Jung and Kun Pyo Lee. 2017. Curiosity or certainty? A qualitative, comparative analysis of Couchsurfing and Airbnb user behaviors. In *Conference on Human Factors in Computing Systems - Proceedings*, 1740–1747. DOI: https://doi.org/10.1145/3027063.3053162

[32] Jiwon Jung, Susik Yoon, Seunghyun Kim, Sangkeun Park, Kun Pyo Lee, and Uichin Lee. 2016. Social or financial goals? Comparative analysis of user behaviors in Couchsurfing and Airbnb. In *Conference on Human Factors in Computing Systems - Proceedings*, 2857–2863. DOI: https://doi.org/10.1145/2851581.2892328

[33] Maximilian Klein, Jinhao Zhao, Jiajun Ni, Isaac Johnson, Benjamin Mako Hill, and Haiyi Zhu. 2017. Quality standards, service orientation, and power in Airbnb and Couchsurfing. *Proc. ACM Human-Computer Interact.* 1, CSCW (2017). DOI: https://doi.org/10.1145/3134693

[34] Boriana Koleva, Jocelyn Spence, Steve Benford, Hyosun Kwon, Holger Schnädelbach, Emily Thorn, William Preston, Adrian Hazzard, Chris Greenhalgh, Matt Adams, Ju Row Farr, Nick Tandavanitj, Alice Angus, and Giles Lane. 2020. Designing Hybrid Gifts. *ACM Trans. Comput. Interact.* 27, 5 (August 2020), 1–33. DOI: https://doi.org/10.1145/3398193

[35] Christopher Koopman, Matthew D. Mitchell, and Adam D. Thierer. 2014. The Sharing Economy and Consumer Protection Regulation: The Case for Policy Change. *J. Bus. Entrep. Law* 8, 2 (2014), 529–545. DOI: https://doi.org/10.2139/ssrn.2535345

[36] Isak Ladegaard. 2021. Strangers in the sheets: How Airbnb hosts overcome uncertainty. *Socio-Economic Rev.* 19, 4 (November 2021), 1245–1264. DOI: https://doi.org/10.1093/ser/mwab013

[37] Airi Lampinen and Coye Cheshire. 2016. Hosting via airbnb: Motivations and financial assurances in monetized network hospitality. In *Conference on Human Factors in Computing Systems - Proceedings*, 1669–1680. DOI: https://doi.org/10.1145/2858036.2858092

[38] J. David Lewis and Andrew Weigert. 1985. Trust as a social reality. *Soc. Forces* 63, 4 (June 1985), 967–985. DOI: https://doi.org/10.1093/sf/63.4.967

[39] Michael W. Macy and John Skvoretz. 1998. The evolution of trust and cooperation between strangers: a computational model. *Am. Sociol. Rev.* 63, 5 (1998), 638–660. DOI: https://doi.org/10.2307/2657332

[40] Hiroaki Masaki, Kengo Shibata, Shui Hoshino, Takahiro Ishihama, Nagayuki Saito, and Koji Yatani. 2020. Exploring Nudge Designs to Help Adolescent SNS Users Avoid Privacy and Safety Threats. In *Conference on Human Factors in Computing Systems - Proceedings*, Association for Computing Machinery. DOI: https://doi.org/10.1145/3313831.3376666

[41] Marcel Mauss. 1925. *The Gift: Forms and Functions of Exchange in Archaic Societies*. The Norton Library, New York.

[42] Karolina Mikołajewska-Zając. 2018. Terms of reference: The moral economy of reputation in a sharing economy platform. *Eur. J. Soc. Theory* 21, 2 (2018), 148–168. DOI: https://doi.org/10.1177/1368431017716287

[43] Makarand Mody, Tarik Dogru, and Courtney Suess-Raeisinafchi. 2017. The hotel industry's Achilles Heel? Quantifying the negative impacts of Airbnb on Boston's hotel performance. *Bost. Hosp. Rev.* 5, 3 (2017), 1–11. Retrieved September 14, 2022 from www.bu.edu/bhr

[44] Linda D. Molm, David R. Schaefer, and Jessica L. Collett. 2009. Fragile and resilient trust: Risk and uncertainty in negotiated and reciprocal exchange. *Sociol. Theory* 27, 1 (March 2009), 1–32. DOI: https://doi.org/10.1111/j.1467-9558.2009.00336.x

[45] Linda D. Molm, Nobuyuki Takahashi, and Gretchen Peterson. 2000. Risk and trust in social exchange: An experimental test of a classical proposition. *Am. J. Sociol.* 10, 5 (2000), 1396–1427. DOI: https://doi.org/10.1086/210434

[46] Jennie Germann Molz. 2012. CouchSurfing and network hospitality: "It's not just about the furniture." *Hospitality and Society 1*, 215–225. DOI: https://doi.org/10.1386/hosp.1.3.215_2

[47] Fayika Farhat Nova, Rashidujjaman Rifat, Pratyasha Saha, Syed Ishtiaque Ahmed, and Shion Guha. 2019. Online sexual harassment over anonymous social media in Bangladesh. In *ICTD '19: Proceedings of the Tenth International Conference on Information and Communication Technologies and Development*, Association for Computing Machinery, 1–12. DOI: https://doi.org/10.1145/3287098.3287107

[48] Giovanni Quattrone, Antonino Nocera, Licia Capra, and Daniele Quercia. 2020. Social Interactions or Business Transactions?What customer reviews disclose about Airbnb marketplace. In *The Web Conference 2020 - Proceedings of the World Wide Web Conference, WWW 2020*, 1526–1536. DOI: https://doi.org/10.1145/3366423.3380225

[49] Alexandrea J. Ravenelle. 2019. *Hustle and Gig: Struggling and Surviving in the Sharing Economy*. University of California Press, Berkeley.

[50] Afsaneh Razi, Ashwaq Alsoubai, Seunghyun Kim, Shiza Ali, Gianluca Stringhini, Munmun De Choudhury, and Pamela J. Wisniewski. 2023. Sliding into My DMs: Detecting Uncomfortable or Unsafe Sexual Risk Experiences within Instagram Direct Messages Grounded in the Perspective of Youth. *Proc. ACM Human-Computer Interact.* 7, CSCW1 (April 2023), 1–29. DOI: https://doi.org/10.1145/3579522

[51] Afsaneh Razi, Seunghyun Kim, Ashwaq Alsoubai, Gianluca Stringhini, Thamar Solorio, Munmun De Choudhury, and Pamela J. Wisniewski. 2021. A Human-Centered Systematic Literature Review of the Computational Approaches for Online Sexual Risk Detection. *Proc. ACM Human-Computer Interact.* 5, CSCW2 (October 2021), 1–38. DOI: https://doi.org/10.1145/3479609

[52] Devan Rosen, Pascale Roy Lafontaine, and Blake Hendrickson. 2011. Couchsurfing: Belonging and trust in a globally cooperative online social network. *New Media Soc.* 13, 6 (March 2011), 981–998. DOI: https://doi.org/10.1177/1461444810390341

[53] Jennifer D. Rubin, Lindsay Blackwell, and Terri D. Conley. 2020. Fragile Masculinity: Men, Gender, and Online Harassment. In *Proceedings of the 2020 CHI Conference on Human Factors in Computing Systems*, Association for Computing Machinery, 1–14. DOI: https://doi.org/10.1145/3313831.3376645

[54] Marshall Sahlins. 1972. *Stone Age Economics*. Aldine Atherton, Chicago. DOI: https://doi.org/10.4324/9780203037416

[55] Juliet B. Schor. 2016. Debating the sharing economy. *J. Self-Governance Manag. Econ.* 4, 3 (2016), 7–22. DOI: https://doi.org/10.22381/jsme4320161

[56] Juliet B. Schor and William Attwood-Charles. 2017. The "sharing" economy: labor, inequality, and social connection on for-profit platforms. *Sociol. Compass* 11, 8 (August 2017). DOI: https://doi.org/10.1111/soc4.12493

[57] Ketaki Shriram and Raz Schwartz. 2017. All are welcome: Using VR ethnography to explore harassment behavior in immersive social virtual reality. In *Proceedings - IEEE Virtual Reality*, IEEE Computer Society, 225–226. DOI: https://doi.org/10.1109/VR.2017.7892258

[58] Jocelyn Spence. 2019. Inalienability: Understanding digital gifts. In *Conference on Human Factors in Computing Systems - Proceedings*, Association for Computing Machinery, 1–12. DOI: https://doi.org/10.1145/3290605.3300887

[59] Vicky Steylaerts and Sean O' Dubhghaill. 2012. Couchsurfing and authenticity: Notes towards an understanding of an emerging phenomenon. *Hosp. Soc.* 1, 3 (2012), 261–278. DOI: https://doi.org/10.1386/HOSP.1.3.261_1

[60] Steven Vallas and Juliet B. Schor. 2020. What do platforms do? Understanding the gig economy. In *Annual Review of Sociology*. 273–294. DOI: https://doi.org/10.1146/annurev-soc-121919-054857

[61] Jessica Vitak, Kalyani Chadha, Linda Steiner, and Zahra Ashktorab. 2017. Identifying women's experiences with and strategies for mitigating negative effects of online harassment. In *Proceedings of the ACM Conference on Computer Supported Cooperative Work, CSCW*, Association for Computing Machinery, 1231–1245. DOI: https://doi.org/10.1145/2998181.2998337

[62] Mark Warner, Juan F. Maestre, Jo Gibbs, Chia Fang Chung, and Ann Blandford. 2019. Signal Appropriation of Explicit HIV Status Disclosure Fields in Sex-Social Apps used by Gay and Bisexual Men. In *Conference on Human Factors in Computing Systems - Proceedings*, Association for Computing Machinery, 15. DOI: https://doi.org/10.1145/3290605.3300922

[63] Karen G. Weiss. 2010. Too ashamed to report: Deconstructing the shame of sexual victimization. *Fem. Criminol.* 5, 3 (2010), 286–310. DOI: https://doi.org/10.1177/1557085110376343

[64] Laura C. Wilson and Katherine E. Miller. 2016. Meta-Analysis of the Prevalence of Unacknowledged Rape. *Trauma, Violence, Abus.* 17, 2 (April 2016), 149–159. DOI: https://doi.org/10.1177/1524838015576391

[65] James D Woods. 1993. *The corporate closet: the professional lives of gay men in America*. Free Press, New York.

[66] Jiang Yang, Mark S. Ackerman, and Lada A. Adamic. 2011. Virtual gifts and guanxi: Supporting social exchange in a Chinese online community. In *Proceedings of the ACM Conference on Computer Supported Cooperative Work, CSCW*, Association for Computing Machinery, 45–54. DOI: https://doi.org/10.1145/1958824.1958832

[67] Tien Ee Dominic Yeo and Yu Leung Ng. 2016. The roles of sensation seeking and gratifications sought in social networking apps use and attendant sexual behaviors. In *ACM International Conference Proceeding Series*, Association for Computing Machinery, 1–9. DOI: https://doi.org/10.1145/2930971.2930990

[68] Jisu Yi, Gao Yuan, and Changsok Yoo. 2020. The effect of the perceived risk on the adoption of the sharing economy in the tourism industry: The case of Airbnb. *Inf. Process. Manag.* 57, 1 (January 2020), 102108. DOI: https://doi.org/10.1016/j.ipm.2019.102108

[69] Viviana A Zelizer. 2000. The purchase of intimacy. *Law Soc. Inq.* 25, 3 (2000), 817–848. DOI: https://doi.org/10.1111/j.1747-4469.2000.tb00162.x

[70] Georgios Zervas, Davide Proserpio, and John Byers. 2021. A First Look at Online Reputation on Airbnb, Where Every Stay is Above Average. *Mark. Lett.* 32, 1 (2021), 1–16. DOI: https://doi.org/10.2139/ssrn.2554500
[71] Julianne Zigos. 2013. Couchsurfing: The best hook-up app ever. *Business Insider*. Retrieved September 14, 2022 from https://www.businessinsider.com/couchsurfing-the-best-hook-up-app-2013-12
[72] Douglas Zytko, Nicholas Furlo, Bailey Carlin, and Matthew Archer. 2021. Computer-Mediated Consent to Sex: The Context of Tinder. *Proc. ACM Conf. Comput. Support. Coop. Work. CSCW* 5, CSCW1 (April 2021), 1–26. DOI: https://doi.org/10.1145/3449288
[73] 2020. We hear you – Couchsurfing Blog. *Couchsurfing*. Retrieved September 1, 2020 from https://blog.couchsurfing.com/we-hear-you/
[74] 2021. Revenue of Airbnb worldwide from 2017 to 2021. *Statista*. Retrieved September 14, 2022 from https://www.statista.com/statistics/1193134/airbnb-revenue-worldwide/
[75] 2021. Couchsurfing's Competitors, Revenue, Number of Employees, Funding, Acquisitions & News - Owler Company Profile. *Owler*. Retrieved December 14, 2021 from https://www.owler.com/company/couchsurfing